\documentclass[fleqn,usenatbib]{mnras}

\usepackage{subcaption}

\usepackage[T1]{fontenc}

\DeclareRobustCommand{\VAN}[3]{#2}
\let\VANthebibliography\thebibliography
\def\thebibliography{\DeclareRobustCommand{\VAN}[3]{##3}\VANthebibliography}

\usepackage{graphicx}	
\usepackage{amsmath}	
\usepackage{amssymb}	

\usepackage{newtxtext,newtxmath}

\title[AB Aur - Possible planet formation through GI]{AB Aurigae: Possible evidence of planet formation through the gravitational instability}

\author[J. Cadman et al.]{\parbox{\textwidth}{James Cadman$^{1,2}$\thanks{E-mail: \texttt{cadman@roe.ac.uk}}, Ken Rice$^{1,2}$, Cassandra Hall$^{3,4,5}$}\vspace{0.4cm}
\\
$^{1}$SUPA, Institute for Astronomy, University of Edinburgh, Blackford Hill, Edinburgh, EH9 3HJ, Scotland, UK\\
$^{2}$Centre for Exoplanet Science, University of Edinburgh, Edinburgh, UK\\
$^{3}$Department of Physics and Astronomy, The University of Georgia, Athens, GA 30602, USA. \\
$^{4}$Center for Simulational Physics, The University of Georgia, Athens, GA 30602, USA.\\ 
$^{5}$Department of Physics and Astronomy, University of Leicester, University Road, Leicester, LE1 7RH, UK\\
}

\date{Accepted 2021 March 26. Received 2021 March 26; in original form 2021 February 12}

\pubyear{2021}

\begin{document}
\label{firstpage}
\pagerange{\pageref{firstpage}--\pageref{lastpage}}
\maketitle

\begin{abstract}
Recent observations of the protoplanetary disc surrounding AB Aurigae have revealed the possible presence of two giant planets in the process of forming. The young measured age of $1-4$\,Myr for this system allows us to place strict time constraints on the formation histories of the observed planets. Hence we may be able to make a crucial distinction between formation through core accretion (CA) or the gravitational instability (GI), as CA formation timescales are typically Myrs whilst formation through GI will occur within the first $\approx10^4-10^5$\,yrs of disc evolution. We focus our analysis on the $4-13$\,M$_{\rm Jup}$ planet observed at $R\approx30$\,AU. We find CA formation timescales for such a massive planet typically exceed the system's age. The planet's high mass and wide orbit may instead be indicative of formation through GI. We use smoothed particle hydrodynamic simulations to determine the system's critical disc mass for fragmentation, finding $M_{\rm d,crit}=0.3$\,M$_{\odot}$. Viscous evolution models of the disc's mass history indicate that it was likely massive enough to exceed $M_{\rm d,crit}$ in the recent past, thus it is possible that a young AB Aurigae disc may have fragmented to form multiple giant gaseous protoplanets. Calculations of the Jeans mass in an AB Aurigae-like disc find that fragments may initially form with masses $1.6-13.3$\,M$_{\rm Jup}$, consistent with the planets which have been observed. We therefore propose that the inferred planets in the disc surrounding AB Aurigae may be evidence of planet formation through GI.
\end{abstract}




\section{Introduction}
Most of the known exoplanets are believed to have formed in discs of gas and dust around young stars. Owing to recent advances in high resolution infrared (IR) imaging we are now capable of observing the planet formation process taking place. Observations of these discs have revealed substructures indicative of the presence of planetary companions, such as rings \citep{alma2015,andrewsetal16,avenhausetal18,bertrangetal18,dipierro2018,dsharp2}, gaps \citep{andrewsetal11,perezetal15,ginskietal16,vanBoekeletal17} and spirals \citep{garufietal13,gradyetal13,bensityetal15,perezetal16,tangetal17,dsharp3,dongetal18}, and recently it has even become possible to directly image giant protoplanets forming \citep{Keppleretal18,Mulleretal18,haffertetal19,boccalettietal20}. Study of these systems may reveal crucial insights into the underlying physics governing the planet formation process.

AB Aurigae is a $2.4\pm0.2$\,M$_{\odot}$, Herbig Ae/Be star \citep{dewarf03}, at a distance $d\approx162.9\pm1.5$\,pc \citep{gaia2018}. Various authors find an age for the star-disc system to be somewhere between $1-4$\,Myr \citep{vandenetal97,dewarf03,pietuetal05}. Measurements of the disc surrounding AB Aurigae find an extended, $R_{\rm out}=400-500$\,AU, low-mass disc, where $M_{\rm d}=0.01$\,M$_{\odot}$, with up to a factor $\sim 10$ uncertainty on the mass estimate \citep{dewarf03,andrewsetal05,corderetal05,semenovetal05}. The stellar accretion rate of $\dot{M}=1.3\times10^{-7}$\,M$_{\odot}$yr$^{-1}$ \citep{salyketal13} is unusually high for a $1-4$\,Myr old system, as the depleted disc mass at this late stage limits the available amount of accretable material.
    
The AB Aurigae disc has been studied extensively owing to its complex substructure, with authors reporting multiple rings \citep{pietuetal05,hashimotoetal11,tangetal12,tangetal17}, bright inner spirals \citep{pietuetal05}, extended CO spirals \citep{tangetal12}, and the possible presence of multiple, planetary-mass companions \citep{pietuetal05,tangetal12,tangetal17}. Recent high resolution, scattered light observations of AB Aurigae performed by \cite{boccalettietal20} using SPHERE provide some of the most spectacular images of a protoplanetary disc to date, revealing detailed spiral features, and placing new constraints on the properties of any potential companions. A kink in the inner spiral at $R\approx30$\,AU is found to be consistent with the presence of a protoplanet with mass of $4-13$\,M$_{\rm Jup}$ (hereafter referred to as planet P1), which is also consistent with conclusions from previous authors \citep{pietuetal05,tangetal12,tangetal17}. The authors also report a point-source located at the outer edge of the inner disc, which is characterised by a gas and dust cavity at $R\approx140$\,AU, for which they tentatively derive a planetary mass of $3$\,M$_{\rm Jup}$ (hereafter referred to as planet P2). Throughout this paper we aim to explore the likely formation history of planet P1.
    
In the core accretion (CA) model of giant planet formation \citep{mizuno80,pollacketal96}, growth proceeds through the steady collisional accumulation of planetesimals onto a rocky core, which may eventually become massive enough for the onset of accretion of a gaseous envelope. Currently this model provides the most popular explanation for the formation of giant planets. However, it has been shown that formation timescales, which may be anywhere up to 10\,Myr, may exceed typical disc lifetimes \citep{haischladalada01}, specifically in the case of giant planets on wide orbits where the planetesimal surface densities will be low. The discovery of systems such as HR 8799, where four ultra wide-orbit ($15$\,${\rm AU} < a < 70$\,${\rm AU}$), super-Jupiter mass ($M_{\rm P} > 5$\,M$_{\rm Jup}$) planets have been directly imaged \citep{maroisetal08,maroisetal10}, is an example of a particularly challenging system to explain through in-situ CA \citep{nerobjorkmann09,kratteretal10}.
    
CA also faces challenges when establishing how the first solids are able to grow up to and beyond metre sizes, as it is anticipated that grains will encounter growth barriers, such as the fragmentation \citep{birnstieletal12}, bouncing \citep{zsometal10} and radial drift \citep{weidenschilling77} barriers. Mechanisms such as the streaming instability \citep{youdingoodman05,youdinjohansen07} may be capable of generating local regions of extremely high particle densities, which may then undergo gravitational collapse to form the first $100-1000$\,km planetesimals \citep{johansenetal07}. Dust trapping in the spiral regions of self-gravitating discs may also provide suitable conditions for accelerated growth \citep{ricelodato04}, and possible fragmentation of the planetesimal disc \citep{riceetal06}. Multi wavelength infrared (IR) observations of discs may allow us to probe their grain size distributions \citep{draine06,williamscieza11,dutreyetal14,testi14,ileeetal2020}, place constraints on the rate of grain growth, and investigate whether significant growth may occur very early in the disc's evolution when it is massive enough to be self-gravitating \citep{dipierro15,cadmanetal20b}.
    
In the gravitational instability (GI) model of planet formation \citep{boss97}, unstable regions of the disc may directly collapse to rapidly form giant gaseous protoplanets and brown dwarfs. In a differentially rotating disc, susceptibility to GI can be determined by considering the Toomre $Q$ parameter \citep{toomre64},
\begin{equation}\label{eq:toomre}
    Q = \frac{c_{\rm s} \Omega}{\mathrm{\pi} \mathrm{G}\Sigma},
\end{equation}
where $c_{\rm s}$ is the local sound speed, $\Omega$ is the orbital frequency in a rotationally supported disc, $\mathrm{G}$ is the gravitational constant and $\Sigma$ is the local surface density. A disc may become susceptible to GI, the growth of spiral substructure, and potentially disc fragmentation, when $Q\lesssim 1.5-1.7$ \citep{durisenetal07}.

Fragmentation of the disc will occur if unstable regions are able to cool at a faster rate than the thermal energy is generated during collapse (i.e. if the cooling rate is greater than the heating rate). If the disc is able to cool rapidly, the instability will continue to grow until the inevitable outcome of fragmentation ensues. This requirement for rapid disc cooling, which can be characterised by a critical cooling rate \citep{gammie01,riceetal05}, demands that fragmentation will occur at large radii where the disc material is less optically thick, hence can cool more efficiently \citep{clarke09,ricearmitage09,halletal2017}.

Calculation of the Jeans mass in a gravitationally unstable disc can be used to estimate the likely initial masses of fragments formed in this way. Using analytic approximations it has been shown that, with some dependence on the level of disc irradiation, fragmentation may initially form objects with masses between a few and a few 10s of Jupiter masses \citep{forganrice11,forganrice13b,cadmanetal20}. Dynamical evolution, migration, tidal stripping and growth will then follow, during which the fragment may contract to form a compact planetary/brown dwarf mass object, or be entirely torn apart and destroyed \citep{nayakshin10a,nayakshin10b,nayakshin11,forganrice13,nayakshinfletcher15,forganetal18,humphriesetal19}.

It has also been shown that discs around higher mass stars ($M_*\geq 2$\,M$_{\odot}$) may be more susceptible to GI \citep{cadmanetal20,haworthetal20}, which is consistent with observations that show a higher occurrence rate of giant planets and brown dwarfs orbiting these systems \citep{johnson07,bowler10,nielsenetal19}. This suggested existence of two distinct populations of exoplanets is indicative of two modes of planet formation.
    
Although disc instability may not be a viable mechanism for directly forming many of the known exoplanets, it may play a role in the early growth of planet building material \citep{ricelodato04}, as multi-fluid simulations of self-gravitating discs have shown significant enhancement of dust abundance present in spiral arms \citep{halletal2020}. It is possible, but still disputed, that it could also lead to the direct formation of the wide-orbit objects that are found via direct imaging \citep{viganetal17}. We find ourselves in a unique position with AB Aurigae, as most of the exoplanets discovered to date have already undergone significant migration and dynamical evolution since their formation. The young age of AB Aurigae places strict time constraints on the possible formation histories of the observed planets, thus it is an ideal site for testing theories of planet formation.
    
In this paper we focus on the formation history of planet P1 (CA vs. GI), and whether it is possible that the AB Aurigae system could be evidence of planet formation through GI. This paper is organised as follows. In Section \ref{sec:CA} we calculate the likely CA formation timescale of planet P1, and in Section \ref{sec:GI} we evaluate the possibility that the planet may have formed directly through GI during AB Aurigae's early evolution. We determine the critical disc-to-star mass ratio for fragmentation in Section \ref{sec:sphmodels}, and use viscous evolution models in Section \ref{sec:GIviscmodels} to predict whether the disc may have ever been massive enough to fragment at some point in the recent past. We place new constraints on the current mass of the disc in Section \ref{sec:abaurmass}, and in Section \ref{sec:GIjeansmass} we calculate the Jeans mass in a gravitationally unstable, AB Aurigae-like disc. We discuss our results and draw conclusions in Sections \ref{sec:discussion} and \ref{sec:conclusion} respectively.

\section{Core accretion}\label{sec:CA}

\subsection{Core accretion timescale}
\subsubsection{Methods}

To model the formation timescale of a gas giant planet through CA, we use a similar approach to that outlined in \cite{idalin04}. We begin by assuming that either a $M_{\rm core,init}=0.01$\,M$_{\oplus}$ or a $M_{\rm core,init}=0.1$\,M$_{\oplus}$ core, with density $\rho_{\rm core}=3.2$\,gcm$^{-2}$, has formed at a semi-major axis, $a$, which we vary between 5\,AU and 50\,AU. For simplicity, we consider planet growth in-situ and neglect any migration through the disc, the effect of which is discussed in Section \ref{sec:discussion}. 

Core growth proceeds at a rate \citep{safronov69}, 
\begin{equation}\label{eq:mdotcore}
    \dot{M}_{\rm core}=\pi R_c^2 \Sigma_{\rm p}\Omega f_g,
\end{equation}
where $R_c$ is the radius of the core, $\Sigma_{\rm p}$ is the local planetesimal surface density, $\Omega$ is the angular frequency and $f_g$ is the gravitational enhancement factor, calculated using the equations from \cite{greenzweiglissauer92}. The local planetesimal surface density, $\Sigma_{\rm p}$, is defined as the surface density of dust within a radial annulus defined by the protoplanet's Hill radius, $R_{\rm H}$, where,
\begin{equation}
    R_H = a\Bigg(\frac{M_{\rm p}}{3M_*}\Bigg)^{1/3},
\end{equation}
where $M_*$ is the mass of the host star and $M_{\rm p}$ is the total planet mass, equal to the sum of the core and envelope masses.

Whilst the core mass is still low, growth initially proceeds through planetesimal accretion, and we update $M_{\rm p}$ using Equation \ref{eq:mdotcore} at each timestep. A planet may begin to retain a gaseous envelope if the core exceeds the critical mass for the onset of gas accretion, $M_{\rm crit}$, where \citep{ikomaetal2000},
\begin{equation}
    M_{\rm crit} = 10\Bigg(\frac{\dot{M}_{\rm core}}{10^{-6}{\rm M_{\oplus}yr^{-1}}}\Bigg)^{0.25}\Bigg(\frac{\kappa}{1{\rm gcm^{-2}}}\Bigg)^{0.25} {\rm M_{\oplus}},
\end{equation}
where $\kappa$ is the planetesimal opacity, for which we use $\kappa=1$\,gcm$^{-2}$.

We use a simple approach to calculate the accretion rate of a gaseous envelope onto the core, $\dot{M}_{\rm gas}$, based on the Kelvin-Helmholtz cooling timescale, $\tau_{\rm KH}$, of the protoplanet, where,
\begin{equation}
    \tau_{\rm KH}=10^9(M_{\rm p}/\mathrm{M}_{\oplus})^{-3}  {\rm years},
\end{equation}
and,
\begin{equation}
    \dot{M}_{\rm gas}=M_{\rm p}/\tau_{\rm KH}.
\end{equation}

This approximation is only valid provided that there is sufficient disc gas present for the planet to accrete, and envelope accretion will cease if the planet is able to deplete all the gas available within its feeding zone. This can be defined in terms of an upper mass limit for in-situ formation, known as the gas isolation mass, $M_{\rm g,iso}$, where,
\begin{equation}
    M_{\rm g,iso} = 50\Bigg(\frac{\Sigma_{\rm g}}{2400 {\rm gcm^{-2}}}\Bigg)^{1.5} \Bigg(\frac{a}{1 \rm AU}\Bigg)^3 \Bigg(\frac{M_*}{\rm M_{\odot}}\Bigg)^{-1/2} {\rm M_{\oplus}},
\end{equation}
where $\Sigma_{\rm g}$ is the local gas density. We prevent further growth once $M_{\rm p} \geq M_{\rm g,iso}$.

We set up the gas component of the disc with a total mass $M_{\rm gas}=0.6$\,M$_{\odot}$, hence a disc-to-star mass ratio, $q=0.25$, and with $\Sigma_{\rm g}\propto R^{-1}$. The surface density profile of the gas disc is evolved using the one dimensional model outlined in \cite{ricearmitage09}, where we assume a radially constant, fixed value for the Shakura-Sunyaev viscous-$\alpha$ of $\alpha=10^{-3}$ \citep{shakurasunyaev73}. The planetesimal component of the disc is set up as,
\begin{equation}
    \Sigma_{\rm p} = f_{\rm dust}\eta_{\rm ice}(R/R_0)^{-1},
\end{equation}
where $f_{\rm dust}$ is a scale factor such that we set $\Sigma_{\rm p}$ at 5\,AU to be $2$\,gcm$^{-2}$, $3$\,gcm$^{-2}$, $5$\,gcm$^{-2}$ and $10$\,gcm$^{-2}$. $\eta_{\rm ice}$ is a constant where,
\begin{equation}
  \eta_{\rm ice} = 
  \begin{cases}
      4.2, & \text{if } a \geq a_{\rm ice} \\
      1, & \text{if } a < a_{\rm ice},
    \end{cases} 
\end{equation}
and $a_{\rm ice}$ is the ice line located at,
\begin{equation}
    a_{\rm ice} = 2.7(M_*/\mathrm{M}_{\odot})^2 {\rm AU}.
\end{equation}
In each case, we allow the planets to evolve in the disc for a maximum of 10\,Myr.

\subsubsection{Results}

\begin{figure*}
    \begin{subfigure}{.45\textwidth}
    \centering
    \includegraphics[width=\linewidth]{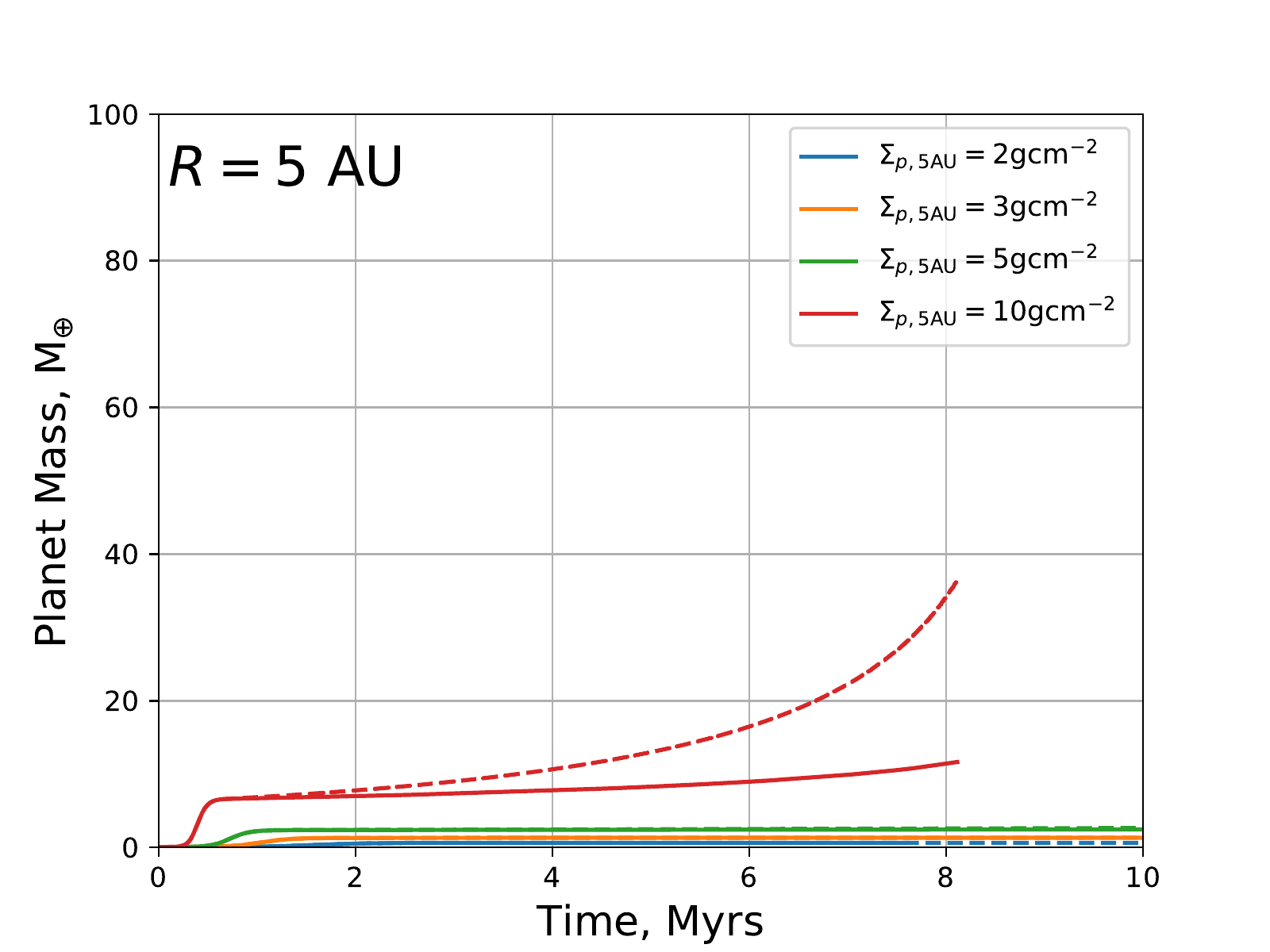}
    \caption{}
    \label{fig:CAa}
    \end{subfigure}
    \begin{subfigure}{.45\textwidth}
    \centering
    \includegraphics[width=\linewidth]{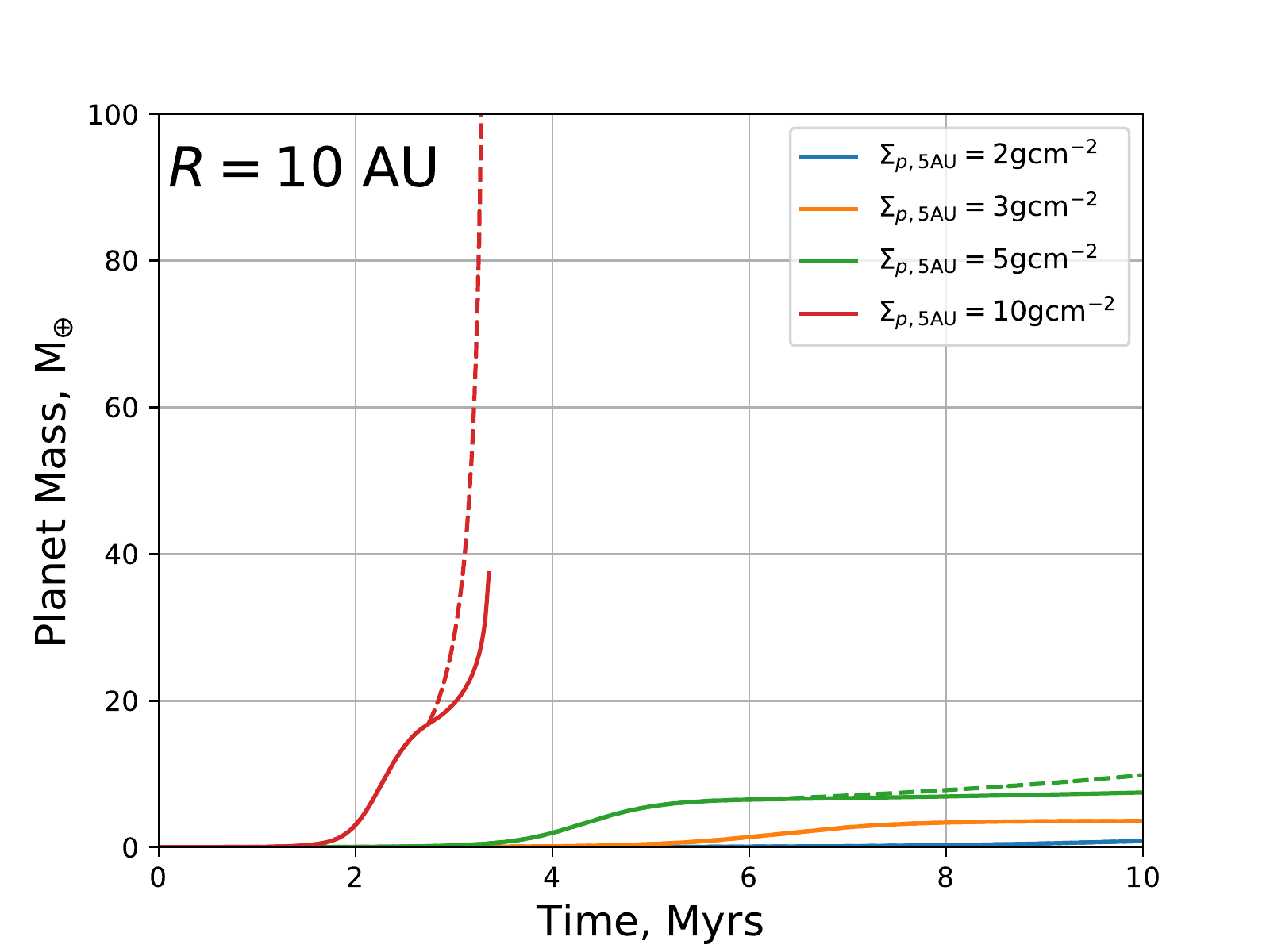}
    \caption{}\label{fig:CAb}
    \end{subfigure}
    \begin{subfigure}{.45\textwidth}
    \centering
    \includegraphics[width=\linewidth]{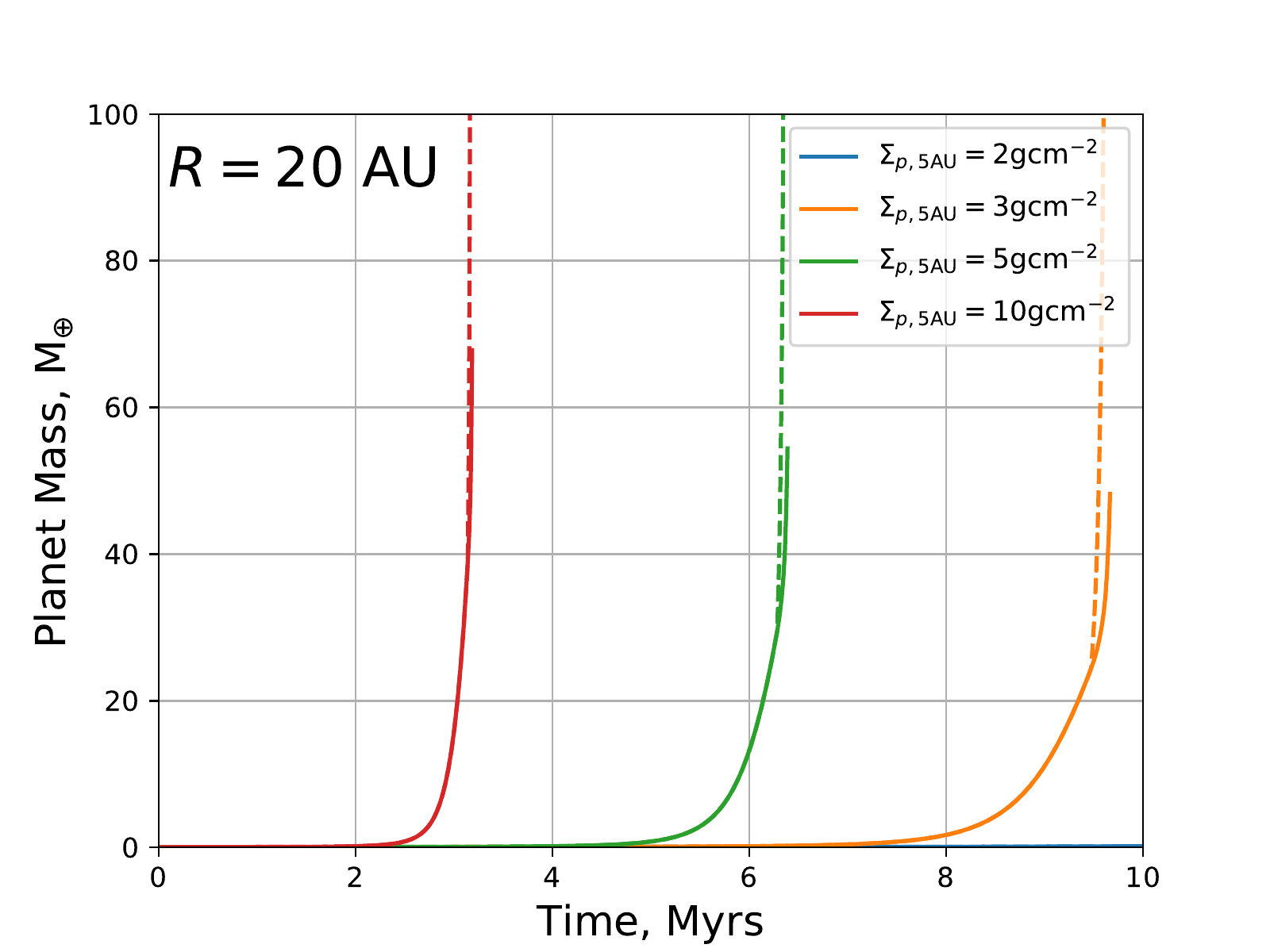}
    \caption{}\label{fig:CAc}
    \end{subfigure}
    \begin{subfigure}{.45\textwidth}
    \centering
    \includegraphics[width=\linewidth]{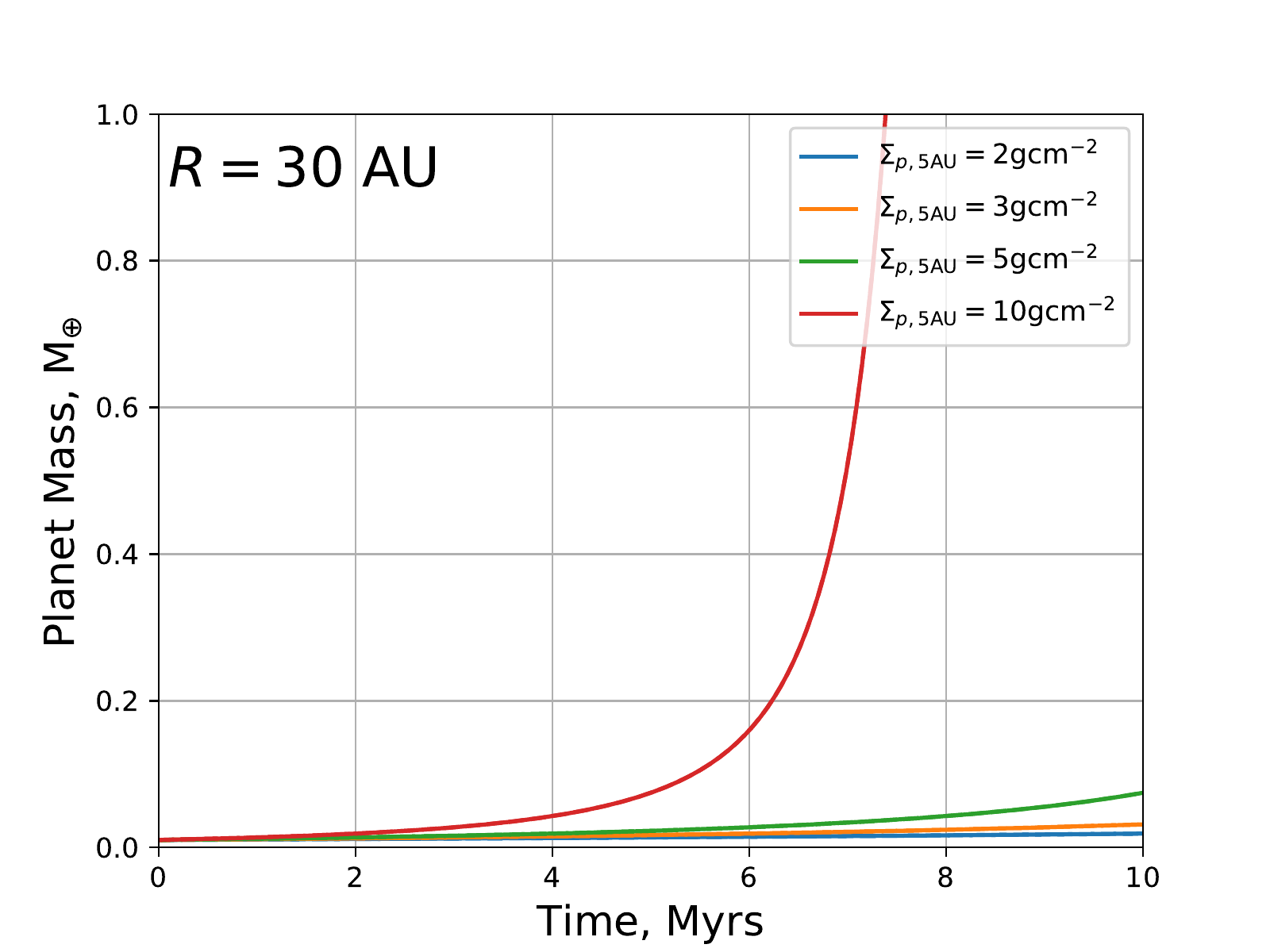}
    \caption{}\label{fig:CAd}
    \end{subfigure}
    \caption{\label{fig:coreacc} Evolution of planet core masses (solid line) and core + envelope masses (dashed lines) for in-situ CA planet formation at radii $R=5$\,AU (\ref{fig:CAa}), $10$\,AU (\ref{fig:CAb}), $20$\,AU (\ref{fig:CAc}) and $30$\,AU (\ref{fig:CAd}) from the stellar host. In each case the models begin with an initial core mass $M_{\rm core,init}=0.01$\,M$_{\oplus}$ at $t=0$. We vary the planetesimal surface densities in the disc such that $\Sigma_{\rm P,5AU}=2$\,gcm$^{-2}$, $3$\,gcm$^{-2}$, $5$\,gcm$^{-2}$ and $10$\,gcm$^{-2}$, which correspond to total planetesimal masses across the disc of $0.012$\,M$_{\odot}$, $0.024$\,M$_{\odot}$, $0.048$\,M$_{\odot}$ and $0.072$\,M$_{\odot}$.}
\end{figure*}

Figure \ref{fig:coreacc} illustrates the resultant planet growth tracks using this formalism, considering the setups with $M_{\rm core,init}=0.01$\,M$_{\oplus}$. Planet formation begins with a phase of core growth, which may either be slow or rapid depending on the local planetesimal surface density. This phase tends to plateau once the local planetesimal surface density is depleted, at which point the planet mass remains approximately constant. The critical core mass for the onset of gas accretion is proportional to the planetesimal accretion rate onto the core, and as the heating from accretion ceases the contraction of a gas envelope may ensue. Wide-orbit, giant planet formation is generally favoured near to, and just beyond the ice line due to the enhancement in the local planetesimal surface density. We calculate $a_{\rm ice}\approx15.6$\,AU for a star of mass $2.4$\,M$_{\odot}$. If the local planetesimal surface density is particularly high, for example near to the ice line in Figure \ref{fig:CAc}, the core mass may pass straight through the critical mass without plateauing. If the local planetesimal surface density is low, for example at a large semi-major axis in Figure \ref{fig:CAd}, the core may never experience significant growth. 

In Table \ref{tab:coreacc} we show the results of 32 runs, where we measure the time for the core to have accreted a significant envelope, of mass equal to the core mass ($M_{\rm p} > 2M_{\rm core}$), and to reach a total planetary mass of $M_{\rm P1}=4$\,M$_{\rm Jup}$, equal to the lower limit of the estimated mass for planet P1. We find it challenging to produce a planet of at least 4\,M$_{\rm Jup}$ in an AB Aurigae-like disc in $\lesssim 1-4$\,Myr. In the majority of setups considered here the planet will either reach its isolation mass before reaching the mass of planet P1, as seen in Figure \ref{fig:CAa}, or will not grow rapidly enough to reach $M_{\rm P1}$ within the duration modelled here. To rapidly form a planet this massive generally requires a significant core has initially formed, in a disc with an extremely high planetesimal surface density, with a planet on a shorter orbit than where planet P1 is currently located.  

The planetesimal surface densities considered here, with $\Sigma_{\rm P,5AU}=2$\,gcm$^{-2}$, $3$\,gcm$^{-2}$, $5$\,gcm$^{-2}$ and $10$\,gcm$^{-2}$ correspond to total planetesimal masses across the disc of $0.012$\,M$_{\odot}$, $0.024$\,M$_{\odot}$, $0.048$\,M$_{\odot}$ and $0.072$\,M$_{\odot}$. These are equivalent to initial dust-to-gas ratios of $0.01$, $0.02$, $0.04$ and $0.06$ if we assume that the dust and gas in the disc decoupled when the gas mass was $1.2$\,M$_{\odot}$ (therefore when $q=0.5$), and that all of this dust then went on to form planetesimals. However, if   any of the dust was depleted by some other mechanism or did not go on to form planetesimals, which would likely be the case, then the planetesimal surface densities used here would demand much higher initial dust-to-gas ratio prior to decoupling. Therefore given the generously high planetesimal surface densities used here, we would consider these core accretion timescales as optimistic lower limits to what we might expect in a realistic disc. 

Mechanisms which may be capable of speeding up these core accretion timescales and have not been included in the models here, such as pebble accretion, are discussed in Section \ref{sec:CAcaveats}.

\begin{table}
    \centering
    \begin{tabular}{c c c c c}
       \hline
        M$_{\rm core,init}$ & $R$ & $\Sigma_{\rm p,5AU}$ & t$_{\rm env}$  & t$_{\rm P1}$\\
        M$_{\oplus}$ & AU & gcm$^{-2}$ & Myr & Myr \\
        (1) & (2) & (3) & (4) & (5) \\
       \hline \hline
       0.01 & 5 & 2 & - & - \\
       0.01 & 5 & 3 & - & - \\
       0.01 & 5 & 5 & - & - \\
       0.01 & 5 & 10 & 6.43 & - \\
       0.01 & 10 & 2 & - & -\\
       0.01 & 10 & 3 & - & -\\
       0.01 & 10 & 5 & - & - \\
       0.01 & 10 & 10 & 3.14 & - \\
       0.01 & 20 & 2 & - & - \\
       0.01 & 20 & 3 & 9.56 & - \\
       0.01 & 20 & 5 & 6.33 & - \\
       0.01 & 20 & 10 & 3.16 & 3.18 \\
       0.01 & 30 & 2 & - & - \\
       0.01 & 30 & 3 & - & - \\
       0.01 & 30 & 5 & - & - \\
       0.01 & 30 & 10 & 8.77 & 8.82 \\
       0.1 & 10 & 2 & - & - \\
       0.1 & 10 & 3 & - & - \\
       0.1 & 10 & 5 & - & - \\
       0.1 & 10 & 10 & 1.88 & - \\
       0.1 & 20 & 2 & 6.81 & - \\
       0.1 & 20 & 3 & 3.95 & 4.05 \\
       0.1 & 20 & 5 & 2.60 & 2.66 \\
       0.1 & 20 & 10 & 1.30 & 1.32 \\
       0.1 & 30 & 2 & - & - \\
       0.1 & 30 & 3 & 9.92 & - \\
       0.1 & 30 & 5 & 6.61 & 6.72 \\
       0.1 & 30 & 10 & 3.33 & 3.38 \\
       0.1 & 50 & 2 & - & - \\
       0.1 & 50 & 3 & - & - \\
       0.1 & 50 & 5 & - & - \\
       0.1 & 50 & 10 & - & - \\
       \hline
    \end{tabular}
    \caption{Results of the core accretion models. (1) Initial core mass. (2) Semi-major axis of core. (3) Planetesimal surface density at 5\,AU. (4) Time before the planet reaches runaway growth, where the envelope mass exceeds the core mass. (5) Time before the planet mass reaches 4\,M$_{\rm Jup}$}
    \label{tab:coreacc}
\end{table}

\section{The Gravitational Instability}\label{sec:GI}

Regions of a protoplanetary disc which become sufficiently gravitationally unstable may undergo a period of rapid collapse to directly form giant gaseous protoplanets and brown dwarfs. GI potentially offers an alternative formation mechanism for wide-orbit, giant planets whose formation timescales are difficult to explain in the CA paradigm. Currently it remains unclear whether GI may be a viable planet formation mechanism; it is uncertain whether discs may ever become sufficiently unstable to fragment, and if they are able to fragment it may be that only stellar and brown dwarf-mass companions are capable of forming in this way. 

\subsection{Critical mass limit for fragmentation}\label{sec:sphmodels}

\subsubsection{Methods}

We use the \textsc{Phantom} \citep{priceetal18} smoothed particle hydrodynamics (SPH) code to determine the critical mass limit for fragmentation in a disc around a 2.4\,M$_{\odot}$ star, analogous to AB Aurigae. SPH allows us to model the detailed hydrodynamics of a fluid, represented as $N$ pseudo-particles, each with an assigned mass, position, velocity and internal energy. A continuous fluid is approximated by calculating particle interactions through a Gaussian kernel function, with a characteristic smoothing length.

We represent the disc with $N=1\times10^6$ SPH particles, distributed between $R_{\rm in}=2.5$\,AU and $R_{\rm out}=400$\,AU with a surface density profile $\Sigma\propto R^{-1}$ and sound speed profile $c_{\rm s}\propto R^{-0.25}$. We modify \textsc{Phantom} such that we model radiative cooling using the hybrid radiative transfer method outlined in \cite{forganetal09}, which combines the polytropic cooling formalism from \cite{stamatellosetal07} and the flux-limited diffusion method \citep{bodenheimer90,cleary99,mayer07}. In using a combination of these two cooling methods we model both the overall energy loss from the system and the detailed energy exchange between neighbouring particles at both high and low optical depths. We assume that disc irradiation leads to a constant background temperature, which we represent as, $T_{\rm irr}=10$\,K. Artificial disc viscosity is modelled using the standard $\alpha-\beta$ viscosity prescription, where we use $\alpha_{\rm SPH}=0.1$ and $\beta_{\rm SPH}=0.2$.

Each disc is allowed to evolve for a maximum of $t=15,550$\,yrs, equal to 3 orbital periods at $R_{\rm out}=400$\,AU, or until fragments form and the computational timestep becomes prohibitively long for the simulations to continue. We calculate the thermalisation timescale, $t_{\rm therm,i}$, from \cite{forganetal09} for each of our disc final states, which represents the time for the disc material to reach thermal equilibrium. We find that in the discs that don't fragment within $15,550$\,yrs, $\max(t_{\rm ttherm,i}) \ll 1$\,kyr. It is therefore reasonable to assume that if these discs have not fragmented after 3 orbital periods, they will not do so in future.

\subsubsection{Results}

\begin{figure*}
\centering
\includegraphics[width=.7\linewidth]{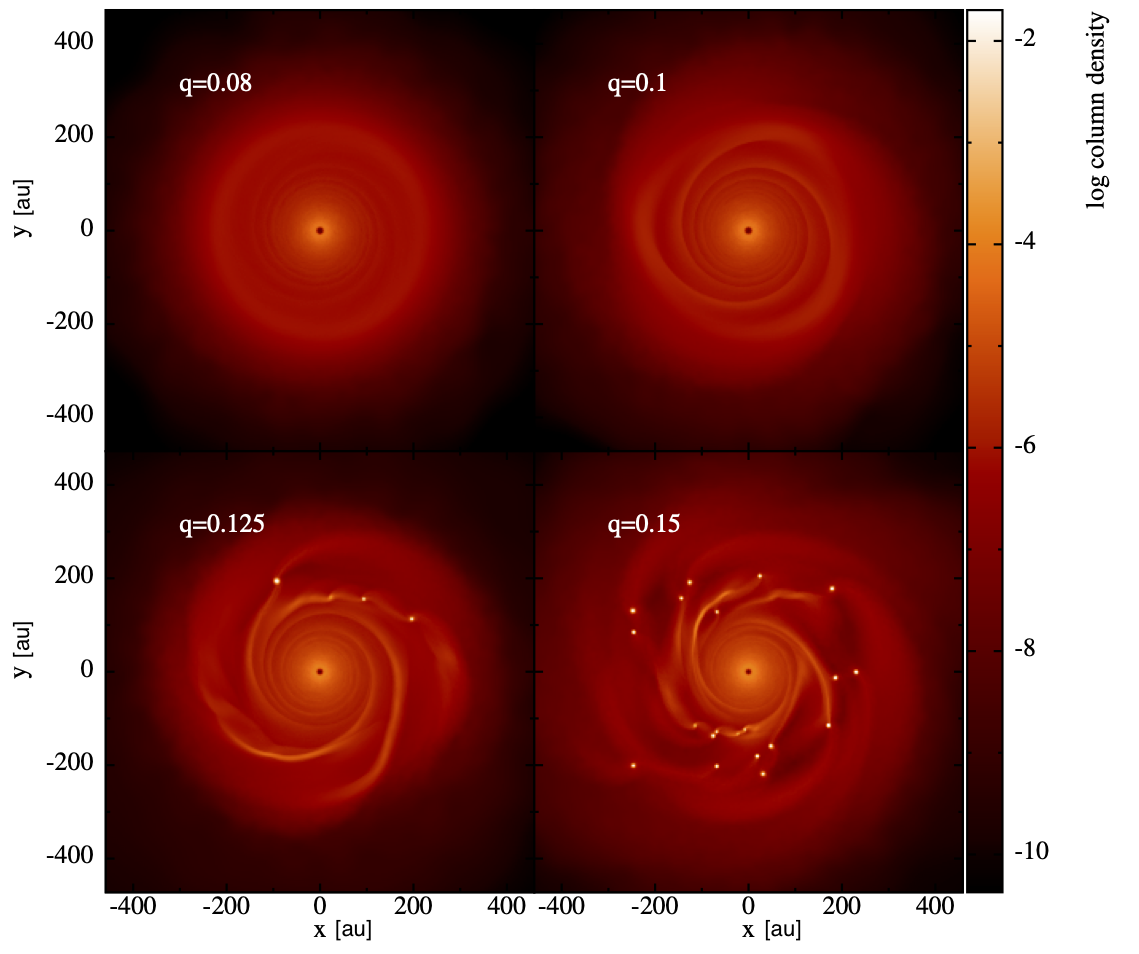}
\caption{\label{fig:sphplots} SPH models of an AB Aurigae-like disc. Each disc is set up with $M_*=2.4$\,M$_{\odot}$, $R_{\rm out}=400$\,AU, $N=1\times10^6$ and $\Sigma \propto R^{-1}$, $c_{\rm s}\propto R^{-0.25}$. We vary the disc-to-star mass ratios within the range $q=0.08-0.15$ ($M_{\rm d}=0.2-0.35$\,M$_{\odot}$). We find the critical disc-to-star mass ratio for fragmentation in an AB Aurigae-like disc to be $q_{\rm crit}=0.125$ ($M_{\rm d,crit}=0.3$\,M$_{\odot}$).}
\end{figure*}

Final states of these SPH simulations are shown in Figure \ref{fig:sphplots}, where we vary the initial disc mass between $0.2$\,M$_{\odot}-0.35$\,M$_{\odot}$, which correspond to disc-to-star mass ratios $q=0.08-0.15$. 

From the final states of these disc models, we expect a disc similar to AB Aurigae to fragment and form multiple clumps if $M_{\rm disc}\geq M_{\rm d,crit}=0.3$\,M$_{\odot} (q\geq0.125)$, and to display non-axisymmetric substructure if $M_{\rm disc}\geq0.25$\,M$_{\odot} (q\geq0.1)$. For $M_{\rm disc}\leq0.2$\,M$_{\odot} (q\leq0.08)$, it is unlikely that the gravitational instability will lead to the growth of significant spirals and, in the absence of a perturber, it should be almost entirely axisymmetric. Therefore given the current low mass state of the AB Aurigae disc we predict that it should be gravitationally stable, as expected.

When also considering a set of discs with outer radii $R_{\rm out}=300$\,AU and $R_{\rm out}=500$\,AU we find that this critical disc-to-star mass ratio has some dependence on disc size, with more extended discs being more stable. When $R_{\rm out}=500$\,AU we find the threshold for fragmentation at $M_{\rm d,crit}=0.35$\,M$_{\odot} (q_{\rm crit}=0.15)$, and when $R_{\rm out}=300$\,AU we find $M_{\rm d,crit}=0.3$\,M$_{\odot} (q_{\rm crit}=0.125)$.

\subsubsection{Subsequent migration of the clumps}

Fragmentation will only occur if the disc is able to radiate energy away at a rate faster than the clump will collapse, hence primarily operates at large radii from the central star where the disc opacity is low thus it can cool efficiently. In the disc with $q=0.125$, the fragment forms at $a\approx200$\,AU, much further out than the current semi-major axis of planet P1. 2D hydrodynamical simulations indicate that once fragments form in a gravitationally unstable disc they will rapidly migrate to the inner regions within a few orbital periods \citep{baruteauetal11}. Computation times become prohibitively long for us to model the long-term migration of clumps in these simulations, as to resolve the high densities at the clump centres requires long integration times. Instead, typical migration of protoplanets can be approximated using the analytic calculations from \cite{nayakshin10a}. For type I migration, the time to move from radii $a_{\rm out}$ to $a_{\rm in}$ will be,
\begin{equation}\label{eq:deltatypeI}
    \Delta t_{\rm mig,I} = \int_{a_{\rm out}}^{a_{\rm in}} \frac{t_{\rm mig,I(a)}}{a} da,
\end{equation}
where,
\begin{equation}\label{eq:typeI}
    t_{\rm mig, I}(a) = \Big(\frac{M_{\rm p}}{M_*}\Omega\Big)^{-1}\frac{H}{a},
\end{equation}
and for type II migration,
\begin{equation}\label{eq:deltatypeII}
    \Delta t_{\rm mig,II} = \int_{a_{\rm out}}^{a_{\rm in}} \frac{t_{\rm mig,II(a)}}{a} da,
\end{equation}
where,
\begin{equation}\label{eq:typeII}
    t_{\rm mig, II}(a) = \frac{1}{\alpha\Omega}\Big(\frac{H}{a}\Big)^{-2},
\end{equation}
where $H$ is the disc scale height at $R=a$. 

Whether a planet is in the type I or type II regime can be established in terms of a transition mass, $M_t$, which roughly corresponds to the mass at which protoplanets become capable of gap-opening. For $M\leq M_t$ (lower-mass, faster migrating protoplanets) the planet will be in the type I regime, and for $M\geq M_t$ (higher-mass, slower migrating protoplanets) the planet will be in the type II regime, where,
\begin{equation}\label{eq:Mtransition}
    M_t = 2M_*\Big(\frac{H}{R}\Big)^3.
\end{equation}

We can calculate the time for planet P1 to migrate from $a_{\rm out}=200$\,AU to $a_{\rm in}=30$\,AU, by substituting $M_*=2.4$\,M$_{\odot}$, $M_{\rm p}=4$\,M$_{\rm Jup}$, $\alpha=0.06$ for a saturated disc, and calculating the azimuthally averaged disc scale height, taken from the SPH disc where $M_{\rm disc}=0.3$\,M$_{\odot}$. Integrating Equations \ref{eq:deltatypeI} and \ref{eq:deltatypeII} we calculate $\Delta t_{\rm mig,I}=6.9$\,kyr and $\Delta t_{\rm mig,II}=1.0$\,Myr. Note that the value of $\alpha$ used here should be considered an upper limit as $\alpha$ will decrease as the planet migrates. Thus the calculated $t_{\rm mig,II}$ would be a lower limit.

From Equation \ref{eq:Mtransition} we calculate the transition mass for gap opening to be $M_t=2.4$\,M$_{\rm Jup}$, which would place planet P1 comfortably in the type II regime. \cite{baruteauetal11} however suggest that GI protoplanets will migrate inwards much faster than the gap opening timescale, and that their migration may be better explained in the type I regime. $\Delta t_{\rm mig,I}$ and $\Delta t_{\rm mig,II}$ are likely more representative of lower and upper limits on the migration timescale of planet P1, and the subsequent migration of a GI protoplanet will be best explained by a combination of both regimes. In either case, these simple calculations demonstrate that, to first approximation, it should be entirely possible for a fragment formed on a wide orbit to migrate inward to the current location of planet P1 within the lifetime of the AB Aurigae disc.

\subsection{Viscous evolution models of AB Aurigae}\label{sec:GIviscmodels}

Despite the system's disc mass being too low to be gravitationally unstable currently, it will likely have been much more massive in the past prior to depletion by stellar accretion and photoevaporative winds, as massive discs will rapidly evolve away from an initially high mass state \citep{halletal19}. Viscous evolution models use analytic prescriptions to calculate the evolution history of a protoplanetary disc's surface density profile. Hence, we may use them to predict the mass evolution history of AB Aurigae. 

\subsubsection{Methods}\label{sec:visagmethods}

Full details of the model used here to calculate the evolution of a disc whose primary source of viscosity is provided by self-gravity can be found in \cite{ricearmitage09}. We also outline the basic equations here. 

Viscous evolution of the surface density, $\Sigma(r,t)$, can be modelled using the one-dimensional prescription from \cite{lyndenbellpringle74,pringle81},
\begin{equation}\label{eq:dsigma/dt}
    \frac{\partial \Sigma}{\partial t}=\frac{3}{r}\frac{\partial}{\partial r}\Big[r^{1/2}\frac{\partial}{\partial r}\Big(\nu\Sigma r^{1/2}\Big)\Big]-\dot{\Sigma}_{\rm wind},
\end{equation}
where $\dot{\Sigma}_{\rm wind}$ represents the photoevaporative mass loss due to radiation from the central star. Here we implement the x-ray photoionization model described in detail in \cite{owenetal11} and assume a moderate x-ray luminosity of $1\times10^{30}$\,erg\,s$^{-1}$, noting that the rate of photoevaporative mass loss scales linearly with x-ray luminosity. Disc viscosity, $\nu(r,t)$, is modelled using the Shakura-Sunyaev viscous$-\alpha$ prescription \citep{shakurasunyaev73},
\begin{equation}\label{eq:viscosity}
    \nu = \alpha c_{\rm s} H,
\end{equation}
where the disc scale height, $H=c_{\rm s}/\Omega$, and $\Omega=\sqrt{GM_*/R^3}$ in a rotationally supported disc. The sound speed, $c_{\rm s}$, is calculated by solving Equation \ref{eq:toomre}, where we force the disc to be in a marginally unstable state with $Q=1.5$.

The volume density can then be calculated as $\rho = \Sigma/2H$, and the temperature, $T$, optical depth, $\tau$, and ratio of specific heats, $\gamma$, can be determined by interpolation of the equation of state table from \cite{stamatellosetal07} using the Rosseland mean opacities from \cite{belllin94}.

To calculate the viscous-$\alpha$ term from Equation \ref{eq:viscosity}, we must first determine the disc cooling time, which requires that we calculate the radiative cooling term \citep{hubney90},
\begin{equation}\label{eq:coolfunc}
    \Lambda = \frac{16\mathrm{\sigma}}{3}(T^4 - T_{\rm irr}^4)\frac{\tau}{1+\tau^2},
\end{equation}
and determine the local cooling time as $t_{\rm cool} = U/\Lambda$, where the energy per unit surface area is, 
\begin{equation}
    U = \frac{c_{\rm s}^2\Sigma}{\gamma(\gamma-1)}.
\end{equation}

In a disc where the primary source of viscosity comes from self-gravity, the effective viscous-$\alpha$ term can be calculated as,
\begin{equation}\label{eq:alpha}
    \alpha = \frac{4}{9 \gamma (\gamma-1)t_{\rm cool}\Omega}.
\end{equation}
We set a lower limit, \textbf{$\alpha_{\rm min}$}, below which we assume that GI is not the dominant source of viscosity but instead, in a sufficiently ionized disc, MRI may dominate, for example. If $\alpha < \alpha_{\rm min}$ we set $\alpha = \alpha_{\rm min}$ and recalculate the disc properties, now no longer requiring the disc to be gravitationally unstable with $Q=1.5$.

Equations \ref{eq:toomre} and \ref{eq:alpha} can then be solved to calculate $c_{\rm s}$ and $\alpha$ for use in Equation \ref{eq:viscosity}, and Equation \ref{eq:dsigma/dt} can be integrated to determine the time evolution of the disc's surface density profile, hence its mass evolution.

\subsubsection{A note on the current mass of the AB Aurigae disc}\label{sec:abaurmass}

Protoplanetary disc masses are notoriously challenging to measure. They often rely on empirical conversions between a disc's flux density and its mass, which requires uncertain assumptions about the disc optical depth, metallicity, dust-to-gas ratio and grain size distribution. Combined with uncertainties in the flux measurement and distance toward the system, mass estimates may be uncertain by up to an order of magnitude, and are usually considered to represent lower bounds. Estimates of the disc mass surrounding AB Aurigae find a low mass disc, with $M_{\rm d}=0.01$\,M$_{\odot}$ and uncertainty up to a factor $\approx 10$ \citep{andrewsetal05,corderetal05,pietuetal05,semenovetal05}.

\begin{figure}
    \centering
    \includegraphics[width=\linewidth]{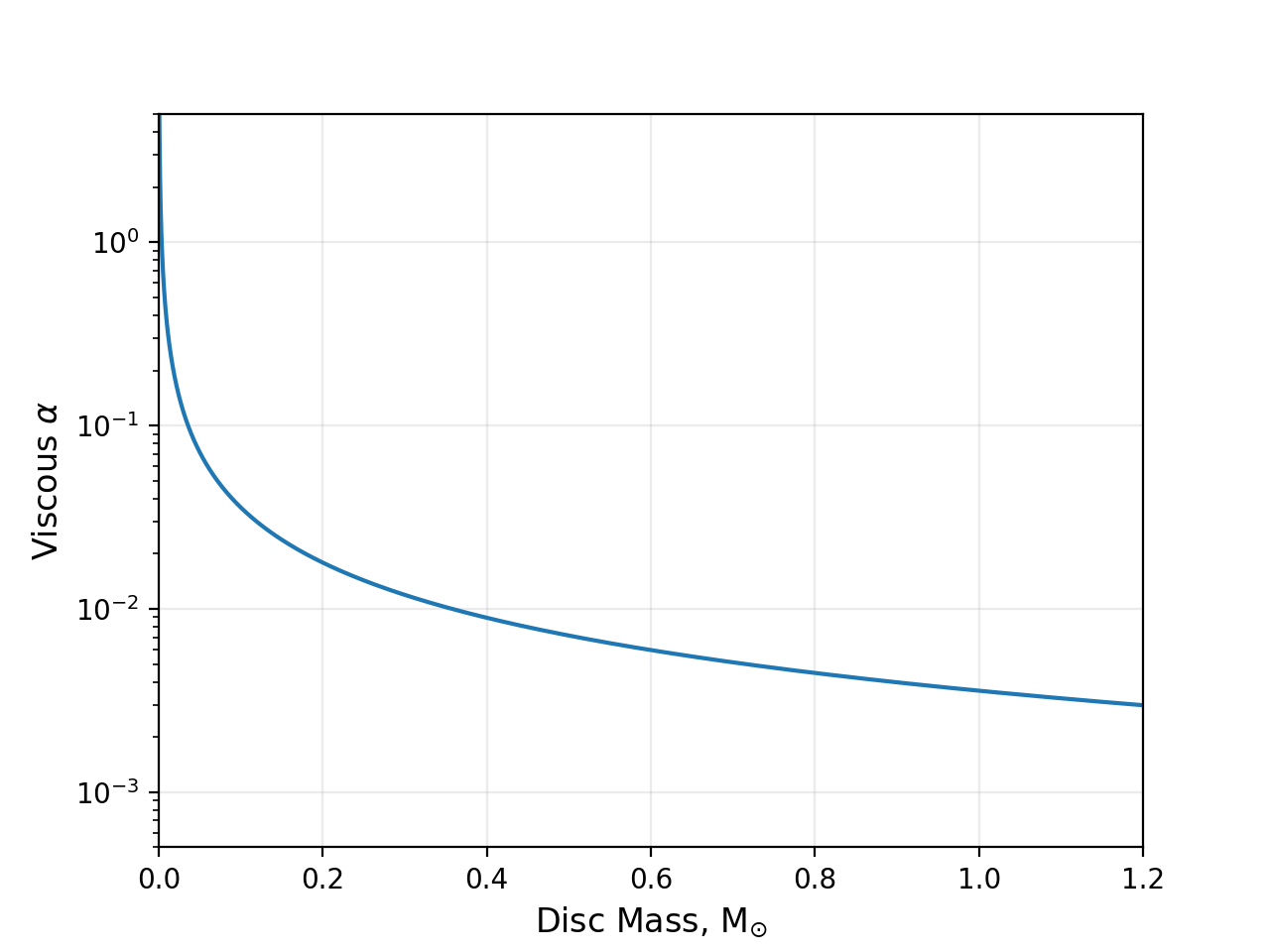}
    \caption{\label{fig:alphaprofile} Viscous$-\alpha$ vs. disc mass for a steady-state disc with $\dot{M}=1.3\times10^{-7}$\,M$_{\odot}$yr$^{-1}$, equal to the mass accretion rate measured in AB Aurigae. We calculate $\alpha$ as a function of disc mass using Equation \ref{eq:alphavsmass}, which assumes that the disc has a radially constant viscous$-\alpha$.}
\end{figure}

The accretion rate onto the star may also provide us a with rough estimate of the disc mass, as it is indicative of the mass reservoir available to the star from the disc. A protoplanetary disc is expected to settle into a steady-state with a constant mass accretion rate \citep{pringle81},
\begin{equation}\label{eq:mdot}
    \dot{M} = \frac{3\pi\alpha c_{\rm s}^2\Sigma}{\Omega} = \rm{constant}.
\end{equation}
In a disc with sound speed profile, $c_{\rm s} = c_{\rm s,0}R^{-0.25}$, and surface density profile, $\Sigma = \Sigma_0 R^{-1}$, the disc may have a radially constant viscous-$\alpha$ given by,
\begin{equation}
    \alpha = \frac{1}{3\pi}\frac{\dot{M}\sqrt{GM_*}}{c_{\rm s,0}^2 \Sigma_0}.
\end{equation}
We can substitute in for $c_{\rm s,0}$ by assuming a flattened disc with $H/R=0.1$ at $R=100$\,AU, and substituting $H=c_{\rm s}/\Omega$, where $H$ is the local disc scale height. Similarly, we can substitute $\Sigma_0$  for the disc outer radius, $R_{\rm out}=400$\,AU and disc mass to obtain an equation in terms of $\alpha$ and $M_{\rm d}$,
\begin{equation}\label{eq:alphavsmass}
    \alpha = \frac{200}{3}\frac{\dot{M}R_{\rm out}}{\sqrt{\mathrm{G}M_*}}\frac{\sqrt{100\rm AU}}{M_{\rm d}}.
\end{equation}
We plot this equation in Figure \ref{fig:alphaprofile} for a star of mass $2.4$\,M$_{\odot}$ and mass accretion rate $\dot{M}=1.3\times10^{-7}$\,M$_{\odot}$yr$^{-1}$ \citep{salyketal13}.

From Figure \ref{fig:alphaprofile} we see that for a very-low mass disc ($M_{\rm d}\leq0.1$\,M$_{\odot}$) to have an accretion rate $\dot{M}=1.3\times10^{-7}$\,M$_{\odot}$yr$^{-1}$ would require a disc viscosity much higher than we would usually expect from a quasi-stable disc, with $\alpha \geq 0.1$. If instead the disc is still massive, with $M_{\rm d}\geq0.1$\,M$_{\odot}$, the viscous-$\alpha$ required to explain the high accretion rate drops significantly. In a quasi-stable disc we might typically expect $\alpha\approx10^{-2}-10^{-4}$ \citep{hartmannetal98,rafikov17}.

We do not attempt to propose an exact disc mass for AB Aurigae here, but instead wish to highlight that in order to explain the system's high accretion rate may require that the disc is more massive than has previously been suggested, and that it is likely at least as massive as the upper bound of the current disc mass estimates.

\begin{table}
    \centering
    \begin{tabular}{|c c|}
         \hline
         $\alpha$ & $M_{\rm disc}$ \\
         & (M$_{\odot}$) \\
         \hline\hline
         \\
         0.1 & $0.04$ \\
         0.01 & $0.36$ \\
         0.001 & $> 1.2$ \\
         \\
         \hline\hline
    \end{tabular}
    \caption{Disc masses corresponding to $\alpha=0.1, 0.01$ and $0.001$ in Figure \ref{fig:alphaprofile}.}
    \label{tab:my_label}
\end{table}

\subsubsection{Results}\label{sec:visagresults}

With this in mind we use these viscous evolution models to predict how long ago the AB Aurigae disc may have been massive enough to exceed the critical mass limit for fragmentation, where $M_{\rm d}=M_{\rm crit}=0.3$\,M$_{\odot}$, assuming the system to have a current disc mass approximately equal to the upper bound on the mass estimate, $M_{\rm d}=0.1$\,M$_{\odot}$. 

In order to do this, we set up discs with initial masses $M_{\rm d}=M_{\rm crit}=0.3$\,M$_{\odot}$ and evolve them forward in time until their mass has been depleted to $M_{\rm d}=0.1$\,M$_{\odot}$, assuming $\alpha_{\rm min}$ values in the range $0.01-0.05$. Discs are set up with initial parameters similar to what we might expect in a young AB Aurigae disc, with $M_*=2.4$\,M$_{\odot}$, $R_{\rm out,init}=400$\,AU, surface density profile $\Sigma\propto R^{-1}$ and temperature profile $T\propto R^{-0.75}$. We assume again that irradiation leads to a constant background temperature $T_{\rm irr}=10$\,K.

\begin{figure}
\centering
\includegraphics[width=\linewidth]{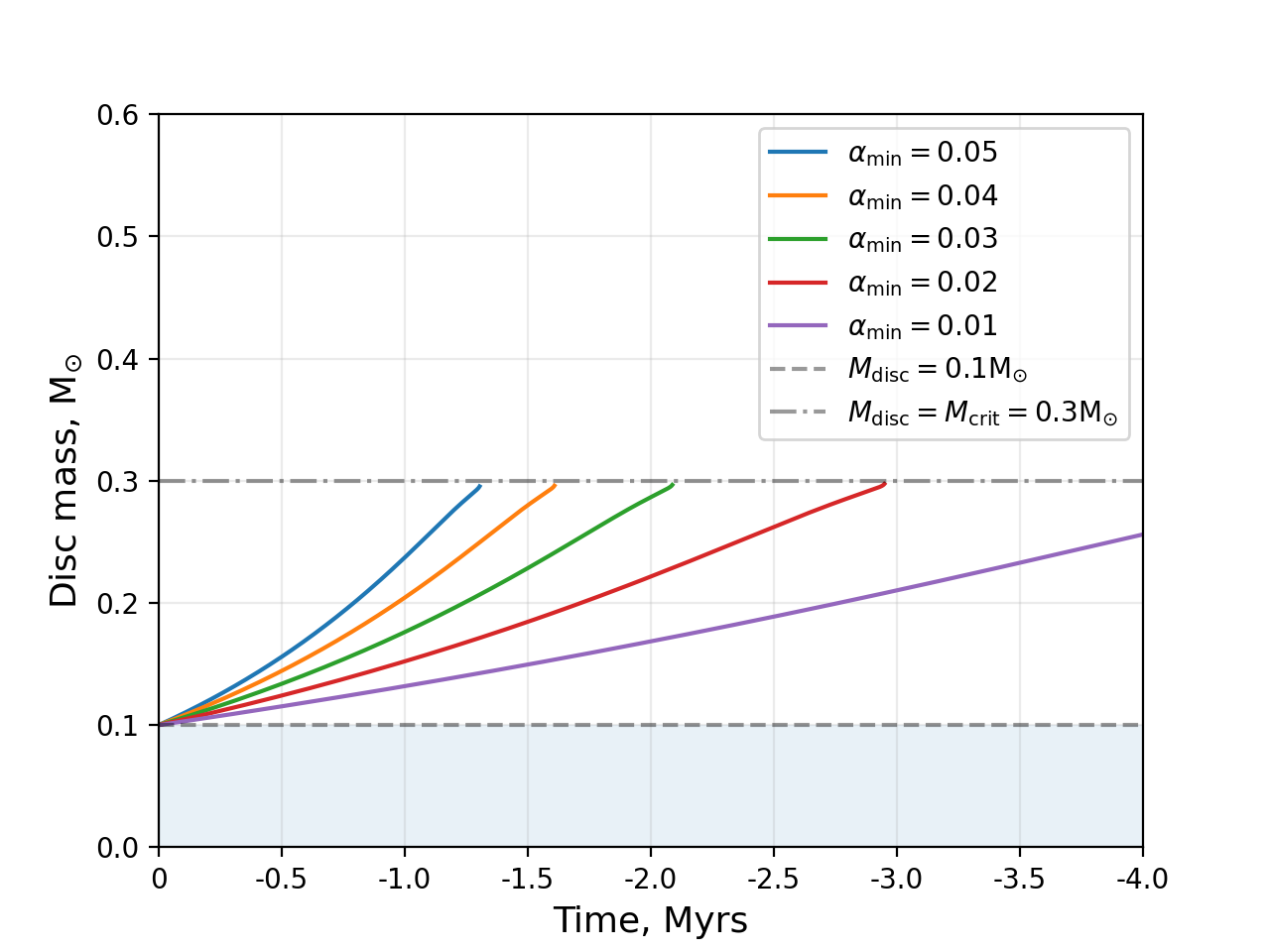}
\caption{\label{fig:viscmodels3} The mass evolution of a disc similar to AB Aurigae, calculated using the viscous evolution models outlined in Section \ref{sec:visagmethods}. The plot begins with a disc mass equal to the current mass of AB Aurigae, with $M_{\rm d}=0.1$\,M$_{\odot}$ at $t=0$, and illustrates how long in the recent past the disc mass may have exceeded the critical mass limit for fragmentation, $M_{\rm d,crit}=0.3$\,M$_{\odot}$. Hence the x-axis measures Myrs in the past. We vary the value of $\alpha_{\rm min}$, which represents a background viscous-$\alpha$ value generated by some process other than disc self-gravity.}
\end{figure}

The results of these models are shown in Figure \ref{fig:viscmodels3}. To illustrate how long in the recent past the AB Aurigae disc may have been massive enough to exceed the fragmentation threshold, we have plotted the disc mass evolution in reverse order. Hence $t=0$ represents the disc in its current state, with $M_{\rm d}=0.1$\,M$_{\odot}$, and the x-axis measures Myrs in the past. For example, in the case of the $\alpha_{\rm min}=0.05$ model, we predict that the AB Aurigae disc may have been more massive than $M_{\rm d,crit}=0.3$\,M$_{\odot}$ approximately $1.3$\,Myrs ago. This approach is equivalent to if we had run the models in Section \ref{sec:visagmethods} backwards, beginning at $M_{\rm d}=0.1$\,M$_{\odot}$.

Discs with higher viscous$-\alpha$ values will evolve at a faster rate, hence the time between the disc mass being in its current state, with $M_{\rm d}=0.1$\,M$_{\odot}$, and exceeding the critical mass limit, $M_{\rm d,crit}=0.3$\,M$_{\odot}$, will be shorter. 

These models again reiterate how it is challenging to reconcile the current low estimated disc mass with the high measured accretion rate, leading us to conclude that AB Aurigae is either currently more massive than observations suggest, or was almost certainly so in its recent past. In the highest accreting case, with $\alpha_{\rm min}=0.05$, we find the accretion rate when $M_{\rm d}=0.1$\,M$_{\odot}$ to be $\dot{M}=4.80\times10^{-8}$\,M$_{\odot}$yr$^{-1}$, and in the lowest accreting case with $\alpha_{\rm min}=0.01$ we find $\dot{M}=9.0\times10^{-8}$\,M$_{\odot}$yr$^{-1}$, both of which are significantly lower than the currently measured value of $\dot{M}=1.3\times10^{-7}$\,M$_{\odot}$yr$^{-1}$ \citep{salyketal13}.

Crucially though, the plots in Figure \ref{fig:viscmodels3} demonstrate how we can trace the AB Aurigae disc back to a previously higher mass state, and how the disc mass may have exceeded the fragmentation threshold in the recent past. When assuming a moderate background $\alpha_{\rm min}$ we find that the disc mass may have exceeded $M_{\rm d,crit}$ within the past $\approx1.25-4$\,Myr. Thus it is plausible that a young AB Aurigae disc may have fragmented to form one or multiple giant gaseous protoplanets during its early evolution.

\subsection{Jeans mass in an AB Aurigae-like disc}\label{sec:GIjeansmass}

The local Jeans mass in a self-gravitating disc can be used, in the case where a region of the disc fragments, to estimate the masses of the bound clumps that will form. \cite{cadmanetal20} derived a revised equation for the Jeans mass in an irradiated self-gravitating disc (presented in its original form in \cite{forganrice13b}) given by,
\begin{equation}\label{eq:mjeans}
    M_{\rm J} = \frac{\sqrt{3}}{32G}\frac{\pi^3 Q^{1/2} c_{\rm s}^2 H}{(1+4.47\sqrt{\alpha})^{1/2}}.
\end{equation}

We can use the same approach as in Section \ref{sec:visagmethods} to calculate how the Jeans mass varies as a function of $\dot{M}$ and $R_{\rm out}$. We assume the disc to be marginally unstable, with $Q=1.5$, and use equation \ref{eq:toomre} to obtain $c_{\rm s}$, and solve equation \ref{eq:alpha} to obtain $\alpha$ for use in equation \ref{eq:mdot}, allowing us to calculate the Jeans mass for a range of disc outer radii and accretion rates.

\begin{figure*}
    \centering
    \includegraphics[width=\textwidth]{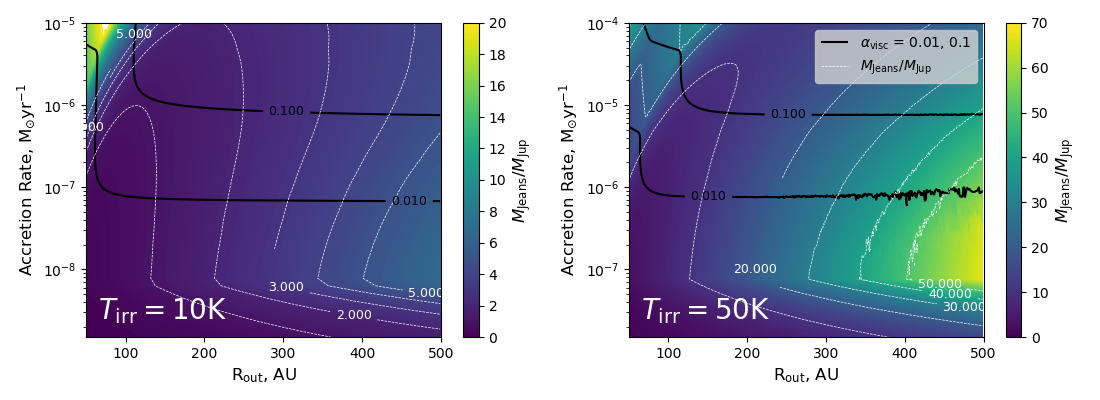}
    \caption{The Jeans mass in a self-gravitating disc surrounding a 2.4\,M$_{\odot}$ star. We consider two cases of disc irradiation, one where it leads to constant background temperature of $T_{\rm irr}=10$\,K (left) and one where it leads to constant background temperature of $T_{\rm irr}=50$\,K (right). We expect a disc to be unstable against fragmentation for $\alpha_{\rm crit}\approx0.06-0.1$, thus we plot contours of $\alpha=0.01$ and $\alpha=0.1$ to indicate regions of parameter space which would likely be unstable against fragmentation. Higher temperatures act to stabilize the disc against the gravitational instability by reducing the effective-$\alpha$. Hence for a given $\dot{M}$ and R$_{\rm out}$ the Jeans masses will be higher when $T_{\rm irr}=50$\,K compared to when $T_{\rm irr}=10$\,K.}
    \label{fig:mjeans}
\end{figure*}

In Figure \ref{fig:mjeans} we plot equation \ref{eq:mjeans}, for a disc around a $2.4$\,M$_{\odot}$ star, with $\dot{M}$ between $1\times10^{-9}$\,M$_{\odot}$yr$^{-1}$ and $1\times10^{-4}$\,M$_{\odot}$yr$^{-1}$, and $R_{\rm out}$ between $50$\,AU and $500$\,AU. We assume that disc irradiation leads to a constant background temperature, and consider two cases where $T_{\rm irr}=10$\,K and $T_{\rm irr}=50$\,K. Higher disc temperatures reduce the effective viscous$-\alpha$ from Equation \ref{eq:alpha} for discs of the same mass, whilst also providing greater pressure support against direct collapse, thus stabilising the system against GI. Hence, for a given $\dot{M}$ and $R_{\rm out}$ the Jeans mass increases as a function of irradiation.

A gravitationally unstable disc may fragment if a collapsing clump is able to cool and radiate energy away at a rate faster than the local dynamical time. This condition can be expressed in terms of a critical value of the dimensionless cooling parameter, $\beta_c = t_{\rm cool}\Omega$, which in turn can be expressed in terms of a critical viscous$-\alpha$ (see equation \ref{eq:alpha}). We typically expect this value to be somewhere between $\alpha_{\rm crit}\approx0.06-0.1$ \citep{gammie01,riceetal05,baehretal17}, thus we include contours of $\alpha=0.01$ and $\alpha=0.1$ in Figure \ref{fig:mjeans} to indicate regions of the parameter space that may fragment.

These plots reiterate that at an earlier stage of AB Aurigae's evolution, when the mass accretion rate was likely higher than it currently is, it is entirely plausible that the disc may have been gravitationally unstable and may have fragmented, as these higher accretion rate states lie in an unstable region of parameter space.

For a given disc radius the minimum Jeans mass doesn't vary much whether we assume fragmentation can only occur for $\alpha\geq0.01$ or $\alpha\geq0.1$. Assuming that $\alpha_{\rm crit}=0.1$ we find the minimum Jeans masses at $R=200$\,AU, $300$\,AU and $400$\,AU to be $1.6$\,M$_{\rm Jup}$, $2.5$\,M$_{\rm Jup}$ and $3.4$\,M$_{\rm Jup}$ respectively when $T_{\rm irr}=10$\,K, and to be $10.3$\,M$_{\rm Jup}$, $12.4$\,M$_{\rm Jup}$ and $13.3$\,M$_{\rm Jup}$ respectively when $T_{\rm irr}=50$\,K, roughly coinciding with what we observe from the mass of planet P1.

\section{Discussion}\label{sec:discussion}

\subsection{Implications for formation through CA}

Significant fine tuning of the model parameters is required in Section \ref{sec:CA} to form planet P1 through CA within the strict time constraint of the system's measured age. To form a planet of $4$\,M$_{\rm Jup}$ within $1-4$\,Myr generally requires a planetesimal surface density much higher than would usually be expected, with a total planetesimal mass across the disc $\geq 0.072$\,M$_{\odot}$ when M$_{\rm core,init}=0.01$\,M$_{\odot}$, and $\geq 0.024$\,M$_{\odot}$ when M$_{\rm core,init}=0.1$\,M$_{\odot}$. When $\Sigma_{\rm p,5AU}=2$\,gcm$^{-2}$, hence with a total planetesimal mass across the disc of $0.012$\,M$_{\odot}$, we generally see very slow planet growth.

It is important to note however that we have only considered a simple formalism for our modelling of CA here, and that processes not included in our models, such as planet migration, pebble accretion and disc instabilities, may be capable of accelerating initial growth. We discuss the effect of these next.

\subsection{Limitations of the CA models}\label{sec:CAcaveats}
Migration allows the planet to sample a wider region of the disc, therefore preventing the local planetesimal surface density becoming depleted as rapidly as when it grows in-situ. When we include core migration in Section \ref{sec:CA} the planets generally grow at a faster rate. However we chose to only consider in-situ formation here, as including migration causes all the cores to migrate to the inner disc ($a \lesssim 3$\,AU) away from the location where we currently find planet P1, and toward the regions of higher planetesimal surface density where they accrete at a faster rate. Some other mechanism, such as planet-planet scattering, would then be required to explain planet P1's subsequent migration out to $a\approx30$\,AU. When modelling in-situ formation at the current semi-major axis of planet P1, we see only slow growth when $M_{\rm core,init}=0.1$\,M$_{\oplus}$, and almost no growth when $M_{\rm core,init}=0.01$\,M$_{\oplus}$.

Instabilities in discs may be capable of generating large over-densities of solids, hence they have been suggested as possible mechanisms for accelerated planetesimal growth and, in extreme cases, fragmentation of the disc solids under their self-gravity. The spiral arms of young, GI discs have been shown to cause strong dust-trapping \citep{ricelodato04}, whilst the gravitational collapse of filaments generated in the streaming instability \citep{youdingoodman05,youdinjohansen07} may form planetesimals of radii $100-1000$\,km \citep{johansenetal07,johansenetal11,johansenetal12}, thus providing a possible mechanism for the initial formation of rocky cores. Whilst refraining from modelling the detailed physics of dust trapping through disc instabilities, we can crudely represent local grain enhancements by simply increasing the total dust-to-gas ratio in the disc, which by default will increase the planetesimal surface density local to the accreting core. We account this by increasing the total planetesimal surface density by up to a factor of 6 in Section \ref{sec:CA}.
    
Mechanisms for accelerated growth and rapid core formation become necessary as CA faces challenges when establishing how the first planetesimals are able to grow beyond metre sizes. The initial stages of growth are believed to be slow, as dust grains may encounter growth barriers beyond metre sizes \citep{braueretal08,mordasinietal10}. It has been shown, as consequence of intrinsic gas-dust drag in the disc, that grains of a critical size will radially migrate and be accreted onto the star within a fraction of the disc lifetime \citep[the radial drift barrier,][]{weidenschilling77}. Further, solids of millimetre to centimetre sizes, with Stokes number close to $1$, are expected to have high relative azimuthal velocities, hence grain-grain collisions may become destructive, resulting in shattering \citep[the fragmentation barrier,][]{birnstieletal12}, or neutral and result in recoiling \citep[the bouncing barrier,][]{zsometal10}, both of which prevent a positive outcome of coagulation. In our model we assume that a core of mass $0.01$\,M$_{\oplus}$ or $0.1$\,M$_{\oplus}$, with $R_{\rm core,init}=1.6\times10^3$\,km and $R_{\rm core,init}=3.5\times10^3$\,km respectively, has already formed at $t=0$, therefore avoiding the detailed physics of this initial phase of core growth. Note that these initial core sizes are consistent with, but slightly larger than, the planetesimals expected to form through direct collapse of the dust disc during the streaming instability \citep{johansenetal07,johansenetal11,johansenetal12}.
    
Possibly most importantly, we note that we do not include a prescription for pebble accretion in our model \citep[for a review see][]{johansenetal17}.  Accretion of millimetre to centimetre sized pebbles onto planetesimal cores may have the potential to generate significantly faster growth rates than the planetesimal-planetesimal accretion we consider here. Pebbles may be abundant in protoplanetary discs, since it is a natural outcome from the fragmentation and bouncing barriers. Pebbles of millimetre-centimetre sizes are coupled to the gas in the disc. The gas component orbits at sub-Keplerian velocities due to the outward gas pressure. The solids, which are orbiting at Keplerian velocities, will experience a drag force which, in a smooth, laminar disc, will cause them to radially drift inward. This migration of pebbles can lead to them being transported to within the path of the growing planetesimal core, constantly replenishing the pebbles within the planetesimal's feeding zone and preventing it from reaching its isolation mass as quickly as they do in Section \ref{sec:CA}. If the planetesimal is gravitationally massive and capable of perturbing the velocities of nearby solids, the pebbles may enter into complex trajectories, orbiting and eventually settling down into its gravitational potential well. If the planetesimal's gravitational cross section exceeds its geometric cross section, pebble accretion may become the dominant growth mechanism. In their review paper \cite{johansenetal17} show that pebble accretion may be capable of resolving many of the timescale problems associated with CA, whilst being able to explain the formation of all planet types.

\subsection{Implications for formation through GI}

In Section \ref{sec:sphmodels} we used SPH simulations to determine the critical mass limit for fragmentation in a disc surrounding a $2.4$\,M$_{\odot}$ star, finding that for a $R_{\rm out}=400$\,AU disc, $M_{\rm d,crit}=0.3$\,M$_{\odot}$ ($q_{\rm crit}=0.125$). Whilst we have mostly focused our discussion on the case of single fragment formation from our SPH simulations, it is also likely that multiple clumps may form in a disc with a mass slightly higher than $M_{\rm d,crit}$ (see Figure \ref{fig:sphplots}). The initial formation of multiple protoplanets may then also provide an explanation for the wider-orbit planet P2 which has also been inferred, located at a distance $a\approx140$\,AU from the parent star \citep{boccalettietal20}. We have refrained from analysing the formation history of planet P2, due to its slightly more tentative detection, choosing instead to focus on planet P1. However it would seem that the formation of a $3$\,M$_{\rm Jup}$ planet at $a\approx140$\,AU may be even more challenging to explain in the CA paradigm than is the case for planet P1, as the gas and dust surface densities in the disc will drop off as $\Sigma\propto R^{-1}$, hence will be exceedingly low at such a large radius. As we see only minimal core growth at $R=30$\,AU in Figure \ref{fig:CAd}, it is likely that growth at $R=140$\,AU would be near-negligible. It may then be the case that in fact planets P1 and P2 represent two survivors from several fragments which could have initially formed.

Despite the AB Aurigae disc being far too low mass to be gravitationally unstable currently, models of the system's viscous evolution in Section \ref{sec:GIviscmodels} suggest that it may have been much more massive when it was younger, potentially exceeding the critical mass limit for fragmentation. It seems reasonable to expect that the disc might have previously fragmented in an extended system such as AB Aurigae, as previous studies suggest that fragmentation is inevitable in GI discs at radii, $R\gtrsim50-100$\,AU \citep{rafikov05,whiteworthstama06,clarke09,forganrice11}. Further, \cite{cadmanetal20,haworthetal20} used hydrodynamic simulations to demonstrate that, whilst lower mass stars may support gravitationally stable massive discs, susceptibility to fragmentation increases as a function of stellar mass, and that discs around higher mass stars ($M_*\geq2$\,M$_{\odot}$) may fragment for relatively low disc-to-star mass ratios. AB Aurigae being an extended disc around a higher mass star therefore seems to be an ideal candidate system to search for surviving products of GI.

If the disc had been able to fragment whilst it was young, it is not necessarily true that the clumps will have survived the $1-4$\,Myr lifetime of the AB Aurigae system. We find that fragments may initially form on wide-orbits with $R\gtrsim200$\,AU, and use analytic calculations to predict initial clump masses $1.6-13.3$\,M$_{\rm Jup}$. However subsequent evolution is inevitable, and the fragments will rapidly migrate through the disc \citep{baruteauetal11}.

In the tidal downsizing hypothesis of planet formation \citep{nayakshin10a,nayakshin10b,nayakshin11} GI embryos will cool and contract as they migrate. Dust sedimentation may lead to the formation of a solid core, potentially of mass comparable to that of a terrestrial planet \citep{boss98}. If the embryo's outer layers contract slowly whilst migration occurs rapidly then tidal stripping from the parent star may occur once the embryo reaches the inner disc, as its physical radius may exceed its Hill sphere \citep{nayakshin10a}. It is possible that many of the initially formed fragments may be entirely destroyed during this tidal downsizing process \citep{nayakshinfletcher15,humphriesetal19}. Population synthesis calculations find this may be the true of $\approx50\%$ of GI protoplanets, with the remaining objects eventually residing at $a\gtrsim20$\,AU \citep{forganrice13}, although when including fragment-fragment scattering this survival fraction may be significantly less \citep{forganetal18}. The initial formation of multiple clumps would then be necessary if any are to survive beyond this early phase of evolution. Accretion of material onto the protoplanets will also occur as they migrate through the disc. \cite{kratteretal10} showed that most GI fragments will grow well beyond the mass limit for Deuterium burning, and that any GI-born planets likely represent the low mass tail of the eventual GI fragment mass distribution. The Jeans mass estimates that we present in Section \ref{sec:GIjeansmass} therefore represent those shortly after collapse only, as dynamical evolution will significantly influence the embryo's eventual mass.

We also tentatively suggest that the previous disc mass estimates ($M_{\rm d}\approx0.01$\,M$_{\odot}$) \citep{dewarf03,andrewsetal05,corderetal05,semenovetal05} appear too low to be consistent with the high stellar accretion rate \citep{salyketal13}, which is indicative of the presence of a large mass reservoir. Assuming the disc to be in a quasi-steady state with a radially constant viscous$-\alpha$ suggests a lower limit for the current disc mass as $M_{\rm d}\gtrsim0.1$\,M$_{\odot}$ (see Fig. \ref{fig:alphaprofile}). This rough lower limit is in fact consistent with the upper bound of the uncertainty on the current disc mass estimates. However even when assuming this slightly higher disc mass, we still find the calculated accretion rates from our viscous evolution models in Section \ref{sec:visagresults} to be significantly lower than the accretion rate measured from the system. On the unusually high stellar accretion rate, \cite{tangetal12} suggest a possible explanation is the presence of an inner disc, characterised by a gas/dust cavity observed at $R\approx100$\,AU, which is being replenished through accretion from the remnant envelope above and below the disc midplane. This would suggest that the measured accretion rate does not represent that of a settled, $R_{\rm out}=400$\,AU disc as we have assumed here, and would allow for the existence of a low mass disc whilst being consistent with a high accretion rate. We only attempt to further highlight this discrepancy between the measured disc mass and accretion rate, and note that the current mass of the disc does not significantly affect the overall conclusions from this paper in regards to the formation history of planet P1.

\section{Conclusions}\label{sec:conclusion}

In this paper we have analysed the possible formation history of the $4-13$\,M$_{\rm Jup}$ planet observed at $a\approx30$\,AU within the protoplanetary disc surrounding AB Aurigae \citep{pietuetal05,tangetal12,tangetal17,boccalettietal20}. The young age of the star-disc system places strict constraints on the CA formation timescale, which we find challenging to explain within $1-4$\,Myr. The planet's high mass and wide-orbit are indicative of a planet which may have instead formed through disc instability in the natal AB Aurigae disc.

The key results are as follows.
\begin{enumerate}
    \item Typical in-situ CA formation timescales for planet P1 exceed the system's measured age. Fine tuning of the model parameters is required in order to form a planet of $4$\,M$_{\rm Jup}$ within $1-4$\,Myr, including significant enhancement of the planetesimal surface density in the disc, and, in most cases, that a large planetesimal core with $M_{\rm core,init}=0.1$\,M$_{\oplus}$ has already formed near to the snow line at $t=0$. At the current semi-major axis of planet P1 ($a=30$\,AU) we find extremely slow in-situ growth due to the low disc surface densities at wide orbits. We do not include a prescription for pebble accretion in our models here, but note that it may be capable of speeding up planet growth.
    \item A disc surrounding a $2.4$\,M$_{\odot}$ star, analogous to young AB Aurigae, would have fragmented if its initial mass exceeded $M_{\rm d,crit}=0.3$\,M$_{\odot}$ ($q_{\rm crit}=0.125$). If the disc mass is slightly higher than $M_{\rm d,crit}$ several fragments may form. Formation of multiple fragments would allow margin for some fragment destruction, which is likely inevitable during their subsequent dynamical evolution of GI protoplanets. 
    \item Viscous evolution models of the AB Aurigae disc suggest that it may have been massive enough to exceed $M_{\rm d,crit}$ during its early evolution whilst the disc was still young and massive. We find that a $0.1$\,M$_{\odot}$ disc may have exceeded $M_{\rm d,crit}=0.3$\,M$_{\odot}$ within the past $\approx1.25-4$\,Myr when considering moderate background viscosity.
    \item Fragments will initially form on wide orbits, where the disc material is cool, and then rapidly migrate inwards. Typical migration timescales of a GI protoplanet which formed at $R\approx200$\,AU within a young AB Aurigae disc are found to be shorter than the current age of the system. We use analytic calculations to determine type I and type II migration timescales, finding that for migration from $R_{\rm out}=200$\,AU to $R_{\rm in}=30$\,AU, $\Delta t_{\rm mig,I}=6.9$\,kyr and $\Delta t_{\rm mig,II}=1.0$\,Myr when considering disc conditions taken from our hydrodynamic simulations.
    \item Calculations of the Jeans mass in a moderately irradiated proto-AB Aurigae disc represent what the initial fragment masses might have been immediately after formation. We find that $M_{\rm J}=1.6-13.3$\,M$_{\rm Jup}$, which is consistent with the masses of the planets P1 and P2 in the AB Aurigae disc.
    \item Although we focus our discussion on the formation history of planet P1, we highlight that planet P2 found at $a\approx140$\,AU with an estimated mass $M_{\rm P2}=3$\,M$_{\rm Jup}$ may be even more challenging to reconcile with formation through CA.
\end{enumerate}

We therefore propose that planets P1 and P2 which have been inferred through scattered light observations of the AB Aurigae disc \citep{boccalettietal20} may stand as evidence of planet formation through GI.

\section*{Acknowledgements}
JC would like to acknowledge funding from a Higgs scholarship provided by the Scottish funding council. CH is a former Winton Fellow and part of this work was supported by Winton Philanthropies / The David and Claudia Harding Foundation.

\section*{Data Availability}

The model data generated in this study will be shared on request to the corresponding author.


\bibliographystyle{mnras}
\bibliography{main} 



\bsp	
\label{lastpage}
\end{document}